\documentstyle[12pt]{article}
\begin{document}
\title{ {\bf The relation between the Toda hierarchy  \\        
        and the KdV hierarchy }  }
\author{ {\bf Yunbo Zeng \hspace{1cm} Runliang Lin \hspace{1cm} Xin Cao
} \\
    {\small {\it Department of Applied Mathematics, Tsinghua University,
        Beijing 100084, China.} }  }  
\date{}
\maketitle

{\bf Abstract } 
Under three relations connecting the field variables of Toda flows
and that of KdV flows, we present three new sequences of 
combination of the equations in the Toda hierarchy which have the KdV 
hierarchy
as a continuous limit.  The relation between the Poisson structures of 
the KdV hierarchy and the Toda hierarchy in continuous limit is also
studied.

\par\ \


\par\ \


\par\ \

\par\ \

It is well known that the KdV-type equations can be obtained as
continuous
limits of suitably chosen discrete integrable systems.  The relation 
between KdV equation and continuous limit of Toda lattice was
first described by Toda and Wadati$^{\cite{TW}}$.  
The limit process in the 
realm of inverse scattering was discussed in \cite{CASE}.  
Kupershmidt$^{\cite{Kup}}$ presented a general setting for the
integrable 
discrete system and studied the limit case in this framework.

In recent years the discussion of relation between a hierarchy of
soliton equations and continuous limit of a hierarchy of integrable 
discrete systems has attracted 
some attention$^{[4-9]}$, since the
relation provides method of deforming the soliton hierarchy.
One of the authors and Rauch-Wojciechowski$^{\cite{YZKVM}}$ first
proposed a 
recombination method to study the relation between a hierarchy of KdV
flows and continuous limit of a hierarchy of Kac-Van Moerbeke flows.
This recombination method was also studied in [6,7]. 
An asymptotic series (in the 
lattice spacing) for field variables of Toda hierarchy in terms of
field variable of KdV hierarchy was proposed in \cite{GTD} in order that
the series be approximate solution to the Toda flows to high accuracy
and
by choosing the initial data of the Toda flows in a canonical way the 
behavior of a
certain Toda flow can mimic the KdV flows.
Then a conjecture on the relation between 
KdV hierarchy and limit of Toda hierarchy was suggested.
The main purpose of this letter is to present clear and conclusive 
statement for the relation between KdV hierarchy and continuous 
limit of Toda hierarchy by using three relations connecting the field
variables of Toda flows and that of KdV flows proposed in
${\cite{GTD}}$. 
We construct three
new sequences of recombination equations in the Toda hierarchy which
have
the KdV hierarchy as continuous limit. Also we present the 
relation between
the Poisson structures of the KdV hierarchy and Toda hierarchy in
continuous 
limit.
The method can be applied to all the relations connecting the field
variables
of Toda flows and that of KdV flows.

First, we briefly describe the Toda hierarchy and the KdV hierarchy 
as presented in [10,11]. Consider the following
discrete isospectral problem,
\begin{equation}
\label{TDEIGEN}
        (E + w + v E^{(-1)}) y =\lambda y,
\end{equation}
where $w=w(n,t)$ and $v=v(n,t)$ depend on integer $n\in {\bf Z}$
and $t \in {\bf R},$ $\lambda$ is the spectral parameter,  
shift operator $E$ is defined as
        $$ (E f) (n) = f(n+1), \quad 
        f^{(k)} (n) =E^{(k)} f (n)=f(n+k), \quad n\in {\bf Z}. $$

The equation in the Toda hierarchy associated with (\ref{TDEIGEN}) can
be 
written as
following Hamiltonian equation$^{\cite{TUTD}}$
\begin{equation}
\label{TDHIERARCHY}
        {\left( \begin{array}{c} w \\ v \end{array} \right)}_{t_m}
        =J K_{m+1}=J \frac{\delta H_{m+1}}{\delta u}, \quad 
m=1,2,\cdots,
\end{equation}
where $\frac{\delta}{\delta u}=
{(\frac{\delta}{\delta w}, \frac{\delta}{\delta v})}^{T},$
Poisson tensor $J,$ Hamiltonian $H_i$ and $K_i$ are defined as
        $$ J\equiv \left( \begin{array}{cc}
                0 & J_{12} \\ J_{21} & 0 \end{array} \right)
        \equiv \left( \begin{array}{cc}
                0 & (1-E) v \\ v(E^{(-1)} -1) & 0 \end{array} \right),
$$
\begin{equation}
        K_i=\frac{\delta H_{i}}{\delta u}=
        \left( \begin{array}{c} -b_i^{(1)} \\ \frac{a_i}{v} \end{array}
                \right), 
        \quad 
        H_i=-\frac{b_{i+1}}{i},  \quad      
                i=0,1,\cdots,
\end{equation}
with $a_0=\frac{1}{2},$ $b_0=0,$ and
\begin{equation}
\label{ab}
        b_{i+1}^{(1)}=w b_i^{(1)} -(a_i^{(1)}+a_i), \quad
        a_{i+1}^{(1)}-a_{i+1} = w(a_i^{(1)}-a_i)+vb_i-v^{(1)}b_i^{(2)},
        \quad  c_i=-vb_i^{(1)}, \quad
        i=0,1,\cdots.
\end{equation}
The equations (\ref{TDHIERARCHY}) have the bi-Hamiltonian formulation
\begin{equation}
\label{TDBIHS}
        G K_{i-1} = J K_i, \quad i=1,2,\cdots,
\end{equation}
        $$ G \equiv \left( \begin{array}{cc}
        vE^{(-1)}-v^{(1)}E & w(1-E)v \\
        v(E^{(-1)}-1)w  &  v(E^{(-1)}-E)v \end{array} \right), $$
where $G$ is the second Poisson tensor. We take $a_1=0$, then the first
four 
$K_i$'s are
\begin{equation}
        K_0 = \left( \begin{array}{c}
        0 \\ \frac{1}{2v} \end{array} \right), \quad
        K_1 = \left( \begin{array}{c}
        1 \\ 0 \end{array} \right), \quad
        K_2 = \left( \begin{array}{c}
        w \\ 1 \end{array} \right), \quad
        K_3 = \left( \begin{array}{c}
        v+v^{(1)}+w^2 \\ w+w^{(-1)} \end{array} \right).
\end{equation}

The Schr\"{o}dinger spectral problem is of the form
\begin{equation}
\label{KDVEIGEN}
        (\partial^2_x+q-\overline\lambda) \overline y = 0.
\end{equation}
which is associated with the following KdV hierarchy$^{\cite{Newell}}$
\begin{equation}
        q_{t_m} = B_0 P_m=B_0 \frac{\delta \overline H_m}{\delta q}, 
        \quad m=1,2,\cdots,
\end{equation}
where the vector field possesses the bi-Hamiltonian formulation with two
Poisson tensors $B_0$ and $B_1$
        $$ B_0  P_{k+1} = B_1 P_k, \quad k=0,1,\cdots, $$
        $$  B_0 = \partial\equiv\partial_x, \quad
        B_1 = \frac{1}{4} \partial^3+q \partial+
                \frac{1}{2} q_x, \quad 
        \overline H_i = \frac{4 \bar b_{i+2}}{2i+1},\quad 
        i = 0,1,\cdots,$$
with $\bar b_0=0,$ $\bar b_1=1,$ and
        $$\bar b_{i+1}=(\frac{1}{4}\partial^2+q-
        \frac{1}{2}\partial^{-1}q_x)\bar b_i,
        \quad i=0,1,\cdots, $$
where $\partial^{-1}\partial=\partial\partial^{-1}=1$. The first three 
$P_k$'s are
\begin{equation}
        P_0 = 2, \quad P_1=q, \quad
        P_2 = \frac{1}{4}(3{q}^2+q_{xx}).
\end{equation}
The well-known KdV equation reads
\begin{equation}
\label{KDV}
        q_{t_2}=\frac{1}{4}{(3 q^2+q_{xx})}_x.
\end{equation}

Consider the Toda hierarchy on a lattice with a small step $h$. 
We interpolate the sequences $(w(n))$ and $(v(n))$ with two smooth 
functions of a 
continuous variable $x$, and relate $w(n)$ and $v(n)$ to $q(x)$ and
$g(x)$ by
\renewcommand{\theequation}{\arabic{equation}a}
\begin{equation}
\label{WVEXPAND}
        w(n)=-2+\frac{1}{2}q(x)h^2+g(x) h^3, \quad
        v(n)=1+\frac{1}{2}q(x) h^2 -g(x) h^3,  
\end{equation}
\renewcommand{\theequation}{\arabic{equation}b}
\addtocounter{equation}{-1}
\begin{equation}
        E^{(k)} w(n) = -2+\frac{1}{2} q(x+k h) h^2+ g(x+k h)h^3, \quad
        E^{(k)} v(n) = 1+\frac{1}{2} q(x+k h) h^2- g(x+k h) h^3,
\end{equation}
\renewcommand{\theequation}{\arabic{equation}}
where $g(x)$ is given by (14).
Also we define 
\begin{equation}
        \lambda = \overline \lambda h^2, \quad 
        y(n) = \overline y(x).
\end{equation}
Then it is easy to see that 
the spectral problem operators in (\ref{TDEIGEN}) 
has the expansion
\begin{equation}
        (E+w+vE^{(-1)} - \lambda) y(n) =
        h^2(\partial^2+q-\overline\lambda) \overline y(x) +O(h^3),
\end{equation}
which implies that the Toda spectral problem goes to the Schr\"{o}dinger
spectral problem in a continuous limit. The expansion (13) doesn't
depend on 
the choice of $g(x)$. Gieseker proposed a way in [5] to choose $g(x)$ 
by requirement that $w(n)$ and
$v(n)$ given by (11) be approximate solution to the Toda flows 
to different accuracy when $q(x)$ satisfies the KdV equation. For
example, 
the first three choices of $g(x)=g_1(x)+g_2(x)h$
are as follows, respectively
        \addtocounter{equation}{1}
        $$ g_1(x)=g_2(x)=0, \eqno{(\arabic{equation}a)} $$
        $$ g_1(x)=\frac{1}{8}q_x(x),\quad g_2(x)=0 
                \eqno{(\arabic{equation}b)} $$
        $$ g_1(x)=\frac{1}{8}q_x(x), \quad g_2(x)=-\frac{1}{32}q^2(x)h.
                \eqno{(\arabic{equation}c)} $$ 
It was examined in {\cite{GTD}} that suitable chosen combination of
first
$K_i$'s goes to first $P_i$'s in continuous limit.  These results
suggest 
that
there exists some relation between the KdV hierarchy and continuous
limit of 
Toda hierarchy.
We will show that under (11) with (14a), (14b) and (14c), respectively, 
a certainly defined combinations
of the equations in the Toda hierarchy have the KdV hierarchy as a 
continuous limit. In order to do so, we need following lemmas.

{\bf Lemma 1. } Under the definition (\ref{WVEXPAND}), we have
\begin{equation}
\label{KIEXPAND}
        K_i\equiv
        \left( \begin{array}{c} -b_i^{(1)} \\ \frac{a_i}{v} \end{array}
                \right)
        =\left( \begin{array}{c} \alpha_i \\ \gamma_i \end{array}
        \right) + O(h), \quad i=0,1,\cdots,
\end{equation}
where the constants $\alpha_i$ and $\gamma_i$ are given by
\begin{equation}
\label{w1}
 \alpha_0 = 0, \quad \alpha_1 = 1, \quad 
           \gamma_0=\frac{1}{2}, \quad \gamma_1=0, \quad
            \alpha_i = (-1)^{(i-1)} C_{2i-2}^{i-1}, 
          \quad
                \gamma_i = (-1)^i C_{2i-2}^{i}, \quad i=2,3,\cdots. 
\end{equation}

{\bf Proof: } 
Under the definition (\ref{WVEXPAND}), it is easy to see from $K_0$ that
         $ \alpha_0 = 0,$ $\gamma_0=\frac{1}{2}.$
Notice the first equation in (\ref{ab}), we have
\begin{equation}
\label{ALPHA}
         \alpha_k = -2 \alpha_{k-1} + 2 \gamma_{k-1}, \quad
k=1,2,\cdots. 
\end{equation}
The identity$^{\cite{TUTD}}$
        $$ \sum_{i=0}^{k} (a_i a_{k-i} + b_i c_{k-i}) = 0, \quad
                \quad k=1,2,\cdots, $$
leads to the equation
\begin{equation}
\label{GAMMA}   
        \gamma_k = \sum_{i=1}^{k-1} ( - \gamma_i \gamma_{k-i} + 
                \alpha_i \alpha_{k-i} ), 
        \quad k=1,2,\cdots.
\end{equation}
Using the equation (\ref{ALPHA}), (\ref{GAMMA}) and the combinational 
identity
        $$ \frac{1}{k} C_{2k-2}^{k-1} = 
                \sum_{i=1}^{k-1}  \frac{1}{i(k-i)} 
                        C_{2i-2}^{i-1} C_{2k-2i-2}^{k-i-1}, 
                \quad  k=1,2,\cdots, $$
we can complete the proof of Lemma 1 by induction.

Under the definition (11) and (14) we have the expansions
\begin{equation}\label{J12}
        J_{12}= h\sum_{i=0}^{\infty}d_ih^i=-h\partial - 
                \frac{1}{2} h^2 \partial^2 - 
                (\frac{1}{6}\partial^3+\frac{1}{2} q \partial +
                \frac{1}{2} q_x) h^3 +O(h^4), 
\end{equation}
$$d_4=-\frac 1{4!}\partial^4-\frac 14\partial^2q+\partial g_1,$$
$$d_i=-\frac 1{i}\partial^i-\frac 1{2(i-2)!}\partial^{i-2}q
+\frac 1{(i-3)!}\partial^{i-3} g_1+\frac 1{(i-4)!}\partial^{i-4} g_2,
\quad i\geq 5,$$

\begin{equation}
        J_{21}= h\sum_{i=0}^{\infty}e_ih^i= -h\partial + 
        \frac{1}{2} h^2\partial^2-
                (\frac{1}{6} \partial^3+\frac{1}{2} q \partial ) h^3 + 
                O(h^4),
\end{equation}
$$e_4=\frac 1{4!}\partial^4+\frac 14q\partial^2+g_1\partial,$$
$$e_i=\frac {(-1)^i}{i}\partial^i+\frac
{(-1)^{i-2}}{2(i-2)!}q\partial^{i-2}
-\frac {(-1)^{i-3}}{(i-3)!}g_1\partial^{i-3}-\frac {(-1)^{i-4}}{(i-4)!}
g_2\partial^{i-4},\quad i\geq 5.$$

We define 
$\widetilde J= \left( \begin{array}{cc} 0 &\widetilde J_{21} 
                        \\\widetilde J_{12} & 0
                        \end{array} \right)$ by requiring that
\begin{equation}\label{ze1}
J\widetilde J=I.
\end{equation}
Then it is found that
$$      \widetilde J_{12}=h^{-1}\sum_{i=0}^{\infty}\widetilde d_ih^i =
 - h^{-1}\partial^{-1} + \frac{1}{2} +
                (\frac{1}{2} q \partial^{-1} -\frac{1}{12} \partial) h +
                O(h^2),$$
\begin{equation}\label{ze2}
        \widetilde J_{21}=h^{-1}\sum_{i=0}^{\infty}\widetilde e_ih^i = 
                -h^{-1}\partial^{-1} -\frac{1}{2} +
                (\frac{1}{2} \partial^{-1} q -\frac{1}{12} \partial ) h
+ 
                O(h^2), 
\end{equation}
where $\widetilde d_i, \widetilde e_i$ are determined by recurrence
formulas
$$\widetilde d_{k}=\partial^{-1}\sum_{i=1}^{k}d_i\widetilde
d_{k-i},\quad
\widetilde e_{k}=\partial^{-1}\sum_{i=1}^{k}e_i\widetilde e_{k-i}.$$
Then it is found that
\begin{equation}
\label{ze3}
\widetilde JJf=f+\eta,
\end{equation}
where function vector $\eta$ comes from the integration. The equation 
(\ref{ze1}) and (\ref{ze3}) implies that $J\eta=0$.
Since the kernel of $J$ is $\{K_0, K_1\}$, we have the following lemma.

{\bf Lemma 2.} The $\widetilde J$ defined by equations (\ref{ze1})
and (\ref{ze2}) satisfies
\begin{equation}
\label{ze4}
J\widetilde J=I,\qquad \widetilde JJf=f+\xi K_1+\delta K_0,
\end{equation}
where $\xi, \delta$ are constants.

{\bf Lemma 3. } Under the definition (\ref{WVEXPAND}), we have the
expansion
\begin{equation}
\label{SDEF}
        S\equiv \frac{1}{4} (\widetilde JG)^2+\widetilde JG =
        \left( \begin{array}{cc} S_{11} & S_{12} \\
                S_{21} & S_{22} \end{array} \right),
\end{equation}
$$ S_{11}=S_{22}=
        (\frac{1}{4}\partial^2+\frac{1}{2}\partial^{-1}q\partial
        +\frac{1}{4}\partial^{-1}q_x)h^2+O(h^3), \quad
        S_{12}=S_{21}=(\frac{1}{2}\partial^{-1}q\partial
        +\frac{1}{4}\partial^{-1}q_x)h^2+O(h^3),  $$

and 
        $$ S_{ij}+S_{kl} = B_0^{-1} B_1 h^2 +O(h^3), $$
where 
$(i,j,k,l) \in \left\{ (1,1,1,2), (1,1,2,1), (1,2,2,2), (2,1,2,2)
\right\},$

{\bf Proof: }
Under the definition (\ref{WVEXPAND}), we have the following expansions
\begin{eqnarray*}
        G_{11} & = & -2h\partial -(\frac{1}{3} \partial^3+q\partial+
                \frac{1}{2}q_x)h^3+ O(h^4), \\
        G_{12} & = & 2h\partial + h^2\partial^2+
                (\frac{1}{3}\partial^3+\frac{1}{2}q\partial+q_x)h^3
                +O(h^4), \\
        G_{21} & = & 2h\partial - h^2\partial^2+
                (\frac{1}{3}\partial^3+\frac{1}{2}q\partial
                -\frac{1}{2}q_x)h^3+O(h^4), \\
        G_{22} & = & -2h\partial
-(\frac{1}{3}\partial^3+2q\partial+q_x)h^3
                +O(h^4). 
\end{eqnarray*}
Set
\begin{equation}
\label{TDEF}
        T\equiv \widetilde JG = \left( \begin{array}{cc} 0 & 
        \widetilde J_{21} \\
\widetilde J_{12}& 0 
                        \end{array} \right)
                \left( \begin{array}{cc} G_{11} & G_{12} \\ G_{21} &
G_{22} 
                        \end{array} \right)
                = \left( \begin{array}{cc} T_{11} & T_{12} \\ T_{21} &
T_{22}
                        \end{array} \right),
\end{equation}
then the operator has the expansion
        $$ T_{11} =  -2 + \frac{1}{2} h^2 q + O(h^3), \quad
        T_{12} = 2 + h\partial +(\frac{1}{2}\partial^2+q)h^2+O(h^3),  $$
        $$ T_{21}=2- h\partial +(\frac{1}{2}\partial^2-
                \frac{1}{2}\partial^{-1}q_x)h^2
                +O(h^3), \quad 
        T_{22} = -2+ \frac{1}{2}h^2\partial^{-1}q\partial +O(h^3). $$
Denote
\begin{equation}
\label{MDEF}
         T^2 = \widetilde JG\widetilde JG
                = \left( \begin{array}{cc} M_{11} & M_{12} \\ 
                        M_{21} & M_{22}
                        \end{array} \right),
\end{equation}
then we get
        $$ M_{11} = 8+(\partial^2-\partial^{-1}q_x)h^2+O(h^3), \quad
        M_{12} =
-8-4h\partial+(-2\partial^2-3q+\partial^{-1}q\partial)h^2
                +O(h^3), $$
        $$ 
        M_{21}=
-8+4h\partial+(-2\partial^2+2q+\partial^{-1}q_x)h^2+O(h^3), 
        \quad
        M_{22} = 8+(\partial^2+\partial^{-1}q_x) h^2 +O(h^3). $$ 
The expansions of (\ref{TDEF}) and (\ref{MDEF}) lead to Lemma 2.

{\bf Lemma 4. } We have
\begin{equation}\label{w2}
        TK_i =\widetilde JG K_i=K_{i+1}+\delta_{i+1} K_0, \quad
        i = 0,1,\cdots,
\end{equation}
where
\begin{equation}
\label{w3}
 \delta_i = -2( \alpha_i+\gamma_i)=
                (-1)^i \frac{2}{i} C_{2i-2}^{i-1}, \quad i=1,2,\cdots. 
\end{equation}
{\bf Proof: }   
The equations (5) and (\ref{ze4}) give rise to
        $$TK_i=\widetilde JG K_i = K_{i+1}+\xi_{i+1}K_1
+\delta_{i+1}K_0. $$ 
It follows from (\ref{KIEXPAND}), (\ref{ALPHA}) and (\ref{TDEF}) that
        $$ \xi_{i+1}=0, \quad
        \delta_{i+1}=-2(\alpha_{i+1}+\gamma_{i+1}), $$
which together with (\ref{w1}) leads to (\ref{w3}).

If we define 
$$W\equiv \frac{1}{4}G\widetilde JG+G=(W_{ij}), \qquad 1\leq i,j \leq
2,$$ 
it follows from the proof of Lemma 2 that
        $$ W_{11}=W_{22}=
        (-\frac{1}{2}q\partial-\frac{1}{4}q_x)h^3+O(h^4), \quad        
        W_{12}=W_{21}=(-\frac{1}{4}\partial^3-\frac{1}{2}q\partial
        -\frac{1}{4}q_x)h^3+O(h^4). $$

Then we arrive following Propositions.

{\bf Proposition 1. } The relation between the Poisson tensors of
the Toda hierarchy and those of the KdV hierarchy is as follows
\begin{equation}
\label{JWEXPAND}
        J = - B_0 \left( \begin{array}{cc} 0 & 1 \\ 1 & 0 
                \end{array} \right) h + O(h^2), \quad
        W_{ij}+W_{kl} = -B_1 h^3 +O(h^4), 
\end{equation}
where 
$(i,j,k,l) \in \left\{ (1,1,1,2), (1,1,2,1), (1,2,2,2), (2,1,2,2)
\right\}.$

{\bf Proposition 2. }  Let
\begin{equation}
\label{CDP1}
        w(n)=-2+\frac{1}{2}   q(x) h^2, \quad
        v(n)=1+\frac{1}{2}   q(x) h^2,
\end{equation}
then
\begin{equation}
\label{VFP1}
        \widetilde P_k \equiv  \sum_{i=0}^{2k} \beta_{k,i} K_i =
        \frac{1}{2}   P_k h^{2k}
        \left( \begin{array}{c} 1 \\ 1 \end{array} \right)
        + O(h^{2k+1}), \quad k \geq 1,
\end{equation}
and
\begin{equation}
\label{EQNP1}
        {\left( \begin{array}{c} w \\ v \end{array} \right)}_{t_k}
        +\frac{1}{h^{2k-1}} J \widetilde P_k =
        \frac{1}{2}(   q_{t_k} - B_0   P_{k} ) h^2
        \left( \begin{array}{c} 1 \\ 1 \end{array} \right)
        + O(h^3), \quad k\geq 1,
\end{equation}
where
        $$ \beta_{k,2k}=(\frac{1}{4})^{k-1}, \quad
        \beta_{k,i}=\frac{1}{4}\beta_{k-1,i-2}+\beta_{k-1,i-1}, \quad
        2\leq i\leq 2k-1, $$
        $$ \beta_{k,1} = \beta_{k-1,0}+
                \frac{1}{4}\sum_{i=0}^{2k-2} \beta_{k-1,i} \delta_{i+1}, 
                \quad \beta_{k,0} = \sum_{i=0}^{2k-2} \beta_{k-1,i} 
                (\frac{1}{2} \delta_{i+1}+\frac{1}{4}\delta_{i+2}), $$
with
        $$ \beta_{1,0} = -2, \quad \beta_{1,1} = 2, \quad
        \beta_{1,2} = 1, \quad \mbox{and} \quad 
        \delta_i=(-1)^i \frac{2}{i} C_{2i-2}^{i-1}, \quad i=1,2,\cdots.
$$

{\bf Proof: }
Under the definition (\ref{CDP1}), it is 
easy to verify the Proposition 2 for $k=1.$
By induction we have
\begin{eqnarray}
\label{TDINDUC}
        &&(\frac{1}{4}(\widetilde JG)^2+\widetilde JG)\widetilde P_{k-1} 
         =   (\frac{1}{4} \widetilde JG +1) 
                        \sum_{i=0}^{2k-2} 
                        \beta_{k-1,i} (K_{i+1}+\delta_{i+1} K_0 )      
                        \nonumber \\
        & = & \frac{1}{4} 
                \sum_{i=0}^{2k-2} \beta_{k-1,i} (K_{i+2}+\delta_{i+2}
K_0 
                        + \delta_{i+1}K_1 +
                        \delta_{i+1} \delta_1 K_0)   
                + \sum_{i=0}^{2k-2} 
                        \beta_{k-1,i} (K_{i+1}+\delta_{i+1} K_0 )      
                \nonumber \\
        & = & \frac{1}{4} \beta_{k-1,2k-2} k_{2k} +
                \sum_{i=2}^{2k-1} 
                (\frac{1}{4} \beta_{k-1,i-2} + \beta_{k-1,i-1} ) K_i 
              +(\beta_{k-1,0}+\frac{1}{4} 
              \sum_{i=0}^{2k-2} \beta_{k-1,i} \delta_{i+1} )K_1
\nonumber\\
          &&    +    \sum_{i=0}^{2k-2} \beta_{k-1,i} 
                (\frac{1}{2}\delta_{i+1}+\frac{1}{4}\delta_{i+2}) K_0
=\widetilde P_k\equiv \sum_{i=0}^{2k} \beta_{k,i} K_i. 
\end{eqnarray}
Using $S$ in (\ref{SDEF}), the following approximation is deduced
\begin{eqnarray}
\label{KDVINDUC}
        &&(\frac{1}{4}(\widetilde JG)^2+\widetilde JG)\widetilde
P_{k-1}  =  
        \left( \begin{array}{cc} S_{11} & S_{12} \\ S_{21} & S_{22}
         \end{array} \right) 
         (\frac 12P_{k-1} h^{2k-2} 
        \left( \begin{array}{c}  1 \\ 1 \end{array} \right) +
O(h^{2k-1})) 
        \nonumber \\
        & = &
        \left( \begin{array}{c} S_{11}+S_{12} \\ S_{21}+S_{22} 
                \end{array} \right) (\frac 12 P_{k-1} h^{2k-2} 
                +O(h^{2k-1})) \nonumber \\
        & = &\frac 12
        B_0^{-1} B_1 P_{k-1} h^{2k} 
        \left( \begin{array}{c}  1 \\ 1 \end{array} \right) +
O(h^{2k+1}) 
      =\frac 12 P_k h^{2k} 
        \left( \begin{array}{c}  1 \\ 1 \end{array} \right) +
O(h^{2k+1}),
\end{eqnarray}
which together with the equation (\ref{TDINDUC}) gives rise to the
equation
(\ref{VFP1}).
The equation (\ref{EQNP1}) can be obtained by combining 
the equation (\ref{JWEXPAND}), (\ref{CDP1}) and (\ref{VFP1}).

For example, the first three $\widetilde P_k$'s are
\begin{eqnarray}
        \widetilde P_1 & = & -2K_0 +2 K_1 + K_2=
        \left( \begin{array}{c} 2+w \\ 1-\frac{1}{v} \end{array}
\right), 
        \nonumber \\
        \widetilde P_2 & = &
        \frac{3}{2} K_0 - K_1 +\frac{3}{2} K_2 +\frac{3}{2} K_3
        +\frac{1}{4} K_4, \nonumber \\
        \widetilde P_3 & = & 
        -\frac{5}{4}K_0+\frac{3}{4}K_1-\frac{5}{8} K_2 +
        \frac{5}{4} K_3+\frac{15}{8}K_4+\frac{5}{8} K_5 +
        \frac{1}{16} K_6.
\end{eqnarray}

{\bf Proposition 3. }  Let
\begin{equation}
\label{CDP2}
        w(n)=-2+\frac{1}{2}   q(x) h^2
        +\frac{1}{8}   q_x (x) h^3 , \quad
        v(n)=1+\frac{1}{2}   q(x) h^2
        -\frac{1}{8}   q_x (x) h^3,
\end{equation}
then we have
\begin{equation}
        Q_k \equiv 
        \left( \begin{array}{c} Q_{k,1} \\ Q_{k,2} \end{array} \right)
        \equiv \sum_{i=0}^{2k-1} \widetilde \beta_{k,i} K_i,
\end{equation}
         $$ Q_{k,1} +   Q_{k,2} =   P_k h^{2k}+O(h^{2k+1}),
         \quad k \geq 2, $$
and
\begin{equation}
\label{EQNP2}
        (w_{t_k}+v_{t_k})+\frac{1}{h^{2k-1}}( J_{12} Q_{k,2}+J_{21}
Q_{k,1})=
        (   q_{t_k} - B_0   P_{k} ) h^2        
        + O(h^3), \quad k\geq 2.
\end{equation}
where
        $$ \widetilde \beta_{k,2k-1}=(\frac{1}{4})^{k-2}, \quad
        \widetilde \beta_{k,i}=
        \frac{1}{4}\widetilde\beta_{k-1,i-2}+\widetilde\beta_{k-1,i-1},
\quad
        2\leq i\leq 2k-2, $$
        $$ \widetilde \beta_{k,1} = \widetilde \beta_{k-1,0}+
                \frac{1}{4}\sum_{i=0}^{2k-3} \widetilde \beta_{k-1,i} 
                \delta_{i+1}, \quad
        \widetilde \beta_{k,0} = \sum_{i=0}^{2k-3} \widetilde
\beta_{k-1,i} 
                (\frac{1}{2} \delta_{i+1}+\frac{1}{4}\delta_{i+2}), $$
with  $\delta_i$ are given in Proposition 2, and
        $$ \widetilde\beta_{2,0} = 4, \quad \widetilde\beta_{2,1} = -2,
\quad
        \widetilde \beta_{2,2} =2, \quad \widetilde \beta_{2,3} =1. $$

{\bf Proof: }
It is also easy to check Proposition 3 for $k=2.$ Apply the operator
$S$ on $Q_{k-1},$ we can get the similar
equation as (\ref{TDINDUC}) ($\widetilde P_{k-1}$ is substituted by 
$Q_{k-1}$ and the superscript of summation is $2k-3$ and $2k-2$ instead
of
$2k-2$ and $2k-1$, respectively). Upon the assumption of Proposition 3 
for $k-1$ 
        $$ Q_{k-1,1} + Q_{k-1,2} = P_{k-1} h^{2k-2}+O(h^{2k-1}), $$
we have
\begin{eqnarray*}
          &Q_{k,1} + Q_{k,2}  =
            (S_{11}+S_{21}) Q_{k-1,1} + ( S_{12}+S_{22}) Q_{k-1,2}  \\
            &=
            B_0^{-1} B_1 P_{k-1} h^{2k} +O(h^{2k+1}) 
             = P_{k} h^{2k} +O(h^{2k+1}). 
\end{eqnarray*}          
The above formulation and (\ref{JWEXPAND}) lead 
to (\ref{EQNP2}).

The first two $Q_k$'s read
\begin{equation}
        Q_2 = 4 K_0 - 2 K_1 + 2 K_2 + K_3, \quad
        Q_3 = -3 K_0+\frac{3}{2}K_1- K_2 +
        \frac{3}{2} K_3+\frac{3}{2}K_4+\frac{1}{4} K_5.
\end{equation}

{\bf Proposition 4. }  Let
\begin{equation}
\label{CDP3}
        w(n)=-2+\frac{1}{2}   q(x) h^2
        +\frac{1}{8}   q_x (x) h^3
        -\frac{1}{32}  q^2(x) h^4, \ \ 
        v(n)=1+\frac{1}{2}   q(x) h^2
        -\frac{1}{8}   q_x (x) h^3
        +\frac{1}{32}  q^2(x) h^4,
\end{equation}
then
\begin{equation}
        R_k \equiv  \sum_{i=0}^{2k-1} \widetilde \beta_{k,i} K_i =
        \frac{1}{2}   P_k h^{2k}
        \left( \begin{array}{c} 1 \\ 1 \end{array} \right)
        + O(h^{2k+1}), \quad k \geq 2,
\end{equation}
and
\begin{equation}
        {\left( \begin{array}{c} w \\ v \end{array} \right)}_{t_k}+
        \frac{1}{h^{2k-1}} J R_k =
        \frac{1}{2}(   q_{t_k} - B_0   P_{k} ) h^2
        \left( \begin{array}{c} 1 \\ 1 \end{array} \right)
        + O(h^3), \quad k\geq 2.
\end{equation}
where  $\widetilde \beta_{k,i}$ are defined in Proposition 3.

Proposition 4 can be proved by the similar method used in the proof of 
Proposition 2.

Remark 1. The $g(x)$ in (11a) is assumed to be a polynomial of $q(x)$
and the derivatives of $q$ and is chosen in such way in [5] that
when $q(x)$ is a solution of KdV flows, the $w(n), v(n)$ are approximate 
solution to the Toda flows to high accuracy. The geometric meaning of 
this operation is not clear.
Comparing the above propositions, we can conclude that the higher
accuracy
is introduced in the definition (\ref{WVEXPAND}), 
the fewer components are needed in the recombination method to recover 
the KdV hierarchy through the limit process. 
For example,  under (11) with (14a), we need combination of 
$K_2, \cdots, K_6$ to have KdV equation (\ref{KDV}) as a continuous
limit,
however, for (11) with (14b), the combination of 
$K_2, \cdots, K_5$ goes to KdV equation (\ref{KDV}) in continuous limit.

Remark 2. The proposition 1 implies that the Poisson structure of KdV 
hierarchy is recovered combining the entries of the matrices of 
the Toda Poisson tensors in continuous limit. However the geometric
meaning 
of such combinations 
of different entries is not clear.

In addition, apart from the commuting vector fields, 
we can also show that
the conserved functionals, the Lax pairs and restricted flows for Toda 
hierarchy go to those for KdV hierarchy in continuous limits. 

\section*{ Acknowledgment }

The work was supported by the National Basic 
Research Project for
Nonlinear Sciences and the Doctorate Dissertation Foundation of 
Tsinghua University. We are grateful
to the anonymous referee for the helpful comments.




\begin{thebibliography}{s99}
\bibitem{TW} M. Toda and M. Wadati, 
{ J. Phys. Soc. Japan},  34 (1973) 18.

\bibitem{CASE} K.M. Case, J. Math. Phys., 14 (1974) 916; 
K.M. Case and M. Kac, J. Math. Phys., 14 (1974) 594.

\bibitem{Kup} B.A. Kupershmidt, 
Discrete Lax equations and 
differential-difference calculus (Ast\'{e}risque 123, 1985, Soc. Math.
de
France).


\bibitem{YZKVM} Yunbo Zeng and  S. Rauch-Wojciechowski,
{J. Phys. A: Math. Gen.},  {28} (1995) 3825.


\bibitem{GTD} D. Gieseker,  
{ Commun. Math. Phys.}, {181} (1996) 587.

\bibitem{MPKVM} C. Morosi and L. Pizzocchero,  
{Commun. Math. Phys.},  {180} (1996) 505.

\bibitem{YZ98} Yunbo Zeng, Acta Mathematicae Applicatae Sinica,
14 (1998) 176.

\bibitem{MPGEN3} C. Morosi and L. Pizzocchero, 
{J. Phys. A: Math. Gen.}, {31} (1998) 2727.

\bibitem{MP98} C. Morosi and L. Pizzocchero,  
{Revews in Math. Phys.},  {10} (1998) 235.

\bibitem{TUTD} Guizhang Tu,  
{J. Phys. A}, {23} (1990) 3903.

\bibitem{Newell} A.C. Newell,  Soliton in mathematics and physics
({SIAM}, Philadelphia, 1985).

\end {thebibliography}
\end{document}